# Localization in Wireless Sensor Networks: A Survey

JerilKuriakose, Sandeep Joshi and V.I. George

***Abstract---*** *Localization is widely used in Wireless Sensor Networks (WSNs) to identify the current location of the sensor nodes. A WSN consist of thousands of nodes that make the installation of GPS on each sensor node expensive and moreover GPS may not provide exact localization results in an indoor environment. Manually configuring location reference on each sensor node is also not possible for dense network. This gives rise to a problem where the sensor nodes must identify its current location without using any special hardware like GPS and without the help of manual configuration. In this paper we review the localization techniques used by wireless sensor nodes to identify their current location.*

## I. INTRODUCTION

Wireless sensor devices have a wide range of application in surveillance, monitoring etc. Most of the devices in wireless sensor network are made up of off-the shelf materials and deployed in the area of surveillance and monitoring. The responsibility of each sensor node is to identify the changes in its particular region or area. The changes are as movement of animals, decrease or increase in temperature, rainfall etc., and these changes are periodically reported to the aggregation point or the central server. The central server or the aggregation server identifies the area with the help of the location reference sent by the sensor node.

Initially during deployment each sensor nodes are given their location reference. This is done either manually or the sensor nodes automatically calculate the distance with the help of GPS devices attached to it. Installing a GPS device or manually calculating the location cannot be possible in the context of large network because of the excessive cost and workforce involved respectively. To overcome this, sensor nodes are made to identify their locations with the help of neighboring nodes. This paper focuses on the localization techniques used by the sensor nodes to identify their location. Several researches are going on in the field of localization to identify the exact location.

*JerilKuriakose, Research Scholar, School of Computing and Information Technology (SCIT), Manipal University Jaipur, Jaipur. E-mail: kuriakosejeril@gmail.com*
*Sandeep Joshi, Associate Professor, School of Computing and Information Technology (SCIT), Manipal University Jaipur, Jaipur – 303007.*
*V.I. George, Director, School of Electrical, Electronics & Communication Engineering (SEEC), Manipal University Jaipur, Jaipur – 303007.*

The location of the nodes plays a significant role in many areas as routing, surveillance and monitoring, military etc. The sensor nodes must know their location reference to carryout Location-based routing (LR) [1 - 4]. To find out the shortest route, the Location Aided Routing (LAR) [5 - 7] makes use of the locality reference of the sensor nodes. In some industries the sensor nodes are used to identify minute changes as pressure, temperature, gas leak etc., and in military, robots are used to detect landmines where in both the cases location information plays a key part.

## II. CONCEPTS AND PROPERTIES OF LOCALIZATION

Localization of a sensor node is carried out with the help of neighboring nodes. Several localization techniques are discussed in this paper. Fig. 1 illustrates the different techniques or methods used to identify the location of the nodes.

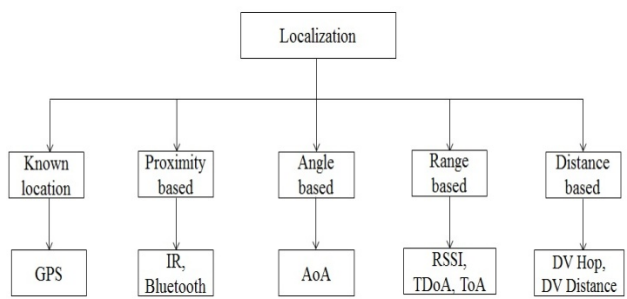

Fig. 1: Overview of Localization

The localization can be classified as known location based localization, proximity based localization, angle based localization, range and distance based localization. In fig. 1 the range and distance based localization are categorized separately, though both are same. For range based localization, special hardware is required to find out the range, however it is not required for distance based localization.

### A. Known location based localization

In this kind of localization the senor nodes know their location in prior. This is done either by manually configuring or using a GPS [8 - 12] device. Manual configuration of the sensor node is done with the help of GPS. The GPS devices are more effective when there are no reference nodes available to get localized. It has a good accuracy with a standard deviation of 4 to 10 meters.

### B. Proximity based localization

In this kind of localization the wireless sensor network is divided into several clusters. Each cluster has a cluster head that has a GPS device. Using Infrared (IR), Bluetooth, etc., the nodes find out the nearness or proximity location.

*C. Angle based localization*

Angle based localization uses the received signals angle or Angle of Arrival [14 - 16] to identify the distance. This method requires special antenna's that are expensive. Because of this reason AOA is mostly used in Base Station's (BS).

*D. Range based localization*

This localization is carried out based on the range. The range is calculated using the Received Signal Strength (RSSI) [17] or Time of Arrival (ToA) [18, 19] or Time Difference of Arrival (TDoA) [13, 20]. In RSSI based localization the receiver sends the signal strength with respect to the sender, and sender calculates the distance based on the signal strength. ToA and TDoA use timing to calculate the range. Time synchronization is an important factor when using ToA and TDoA.

*E. Distance based localization*

Distance based localization technique uses hop distance among each node to localize the node. It uses DV-hop propagation method [21, 22] or DV-distance [22] propagation method for localization.

III. LOCALIZATION TECHNIQUES

The localization techniques can be grouped into two types namely range based and range free approach. Fig. 2 shows the localization techniques grouped into different types.

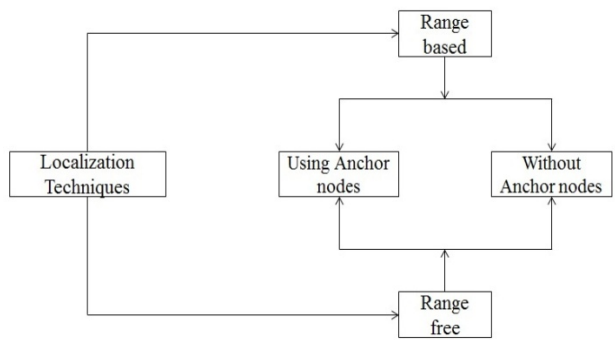

Fig. 2: Categorizing Localization Techniques

*A. Range based approach*

This method uses the range information to calculate the distance between each node. The localization can be carried out with or without the anchor nodes.

*i. Using anchor nodes*

While deploying the sensor network, few are manually configured their location reference either manually or using GPS. These nodes act as the anchor nodes. Other nodes localize themselves with the support of anchor nodes.

Localization is carried out using the range or angle based techniques discussed in the previous section. Each sensor node must be equipped with special hardware to achieve localization. The nodes that use ToA for localization must be time synchronized. The transmitted and received time are used by the sender or receiver to calculate the distance. Fig. 3 shows the range estimation using ToA. The distance between the sender and receiver are calculated as follows: [19]

$$d_{xy} = \frac{1}{2}[(T^x_{recv} - T^x_{trans}) - (T^y_{recv} - T^y_{trans})]$$

where,

$d_{xy}$ is the distance between node X and node Y,

$T^x_{recv}$ is the received power of node X,

$T^x_{trans}$ is the transmitted power of node X,

$T^y_{recv}$ is the received power of node Y,

$T^y_{trans}$ is the transmitted power of node Y.

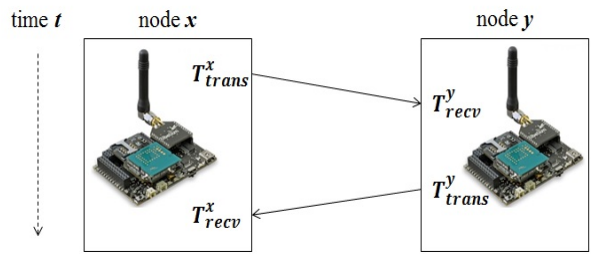

Fig. 3: Range Estimation using ToA

Once the distance is discovered, multilateration is implemented to find out the location reference of the node. RF signal travel at the speed of light, this make the RF propagation to get varied in indoor environments. This made a high localization overhead. In order to overcome the RF propagation in indoor environment, in [13], a combination of RF signals with Ultrasound was proposed. The speed of Ultrasound is lesser when compared to the speed of light. Based on the TDoA of the two signals the distance is calculated. Another method for locating a node using TDoA is done by observing the time for a signal to reach two or more receivers. It is made sure that all the receiver nodes are time synchronized. The TDoA is calculated as follows: [23]

$$\tau = (r_2 - r_1)/c$$

where,

$\tau$ is the TDoA,

$r_2 \& r_1$ are the range from the transmitter to the two receivers,

$c$ is the speed of propagation.

*ii. Without using anchor nodes*

A device that has GPS attached need not require a support from anchor nodes for localization. Triangulation [24] technique is used in GPS to identify the location of the node. The assistance of satellites is required for finding out the location of the sensor node that has GPS device.

*B. Range free approach*

There are few localization techniques that do not require special hardware for localization. They compute their distance based on DV hop or DV distance. The range free approach can be broadly classified into two types as follows,

*i. Using anchor nodes*

Techniques, namely Probability Grid [21] and Kcdlocation [24] works on DV based distance localization. In these techniques few nodes act as anchor nodes, which in turn are used by other nodes for localizing themselves.

*ii. Without using anchor nodes*

Convex Position Estimation technique [28] works without an anchor node. The network is modeled by a central sever giving equations for revealing the distance between the nodes. It uses a good optimization technique to find out the location of the nodes based on the equations.

## IV. PERFORMANCE OF LOCALIZATION SCHEMES

Table 1 shows the performance comparison of different localization schemes. Each localization techniques serve different purposes. More the number of anchor nodes less the localization error. In dense environment the location error tends to increase. This can be controlled by making the network dense.

Table 1: Comparison of Localization Techniques

| Localization Techniques used | Accuracy |
|---|---|
| GPS | 2 to 15 meters |
| Angle based approach | 1 to 6 meters |
| Range based approach | 4 to 8 meters |
| DV based approach | 10 to 20 meters |

## V. CONCLUSION

This paper covered the different localization techniques used and their problems. The scalability of range free approach is more when compared with to range based approach. The localization techniques help by reducing the deployment cost of wireless sensor networks. Currently, there is a trade-off between the localization accuracy and algorithm runtime. Many security and energy issues related to localization that can be considered for future work.